\documentclass[twocolumn, prl]{revtex4}
\usepackage{graphicx}
\usepackage{epsfig}
\usepackage{color}
\usepackage{amsmath}
\begin{document}
\title{Concept of contact spectrum and its applications in atomic quantum Hall states}
\author{Mingyuan He$^*$, Shaoliang Zhang$^*$, Hon Ming Chan,  Qi Zhou}
\affiliation{Department of Physics, The Chinese University of Hong Kong, Shatin, New Territories, HK}
\date{\today}
\begin{abstract}
A unique feature of ultracold atoms is the separation of length scales,
$r_0\ll k_F^{-1}$, where $k_F$ and $r_0$ are the Fermi momentum characterizing
the average particle distance and the range of interaction between atoms
respectively. For $s$-wave scattering, Shina Tan discovered that such
diluteness leads to universal relations, all of which are governed by contact,  among a wide range of thermodynamic quantities.
Here, we show that the concept of contact can be generalized to an arbitrary
partial-wave scattering. Contact of all partial-wave scatterings form a contact
spectrum, which establishes universal thermodynamic relations with notable
differences from those in the presence of $s$-wave scattering alone. Moreover, such a
contact spectrum  has an interesting connection
with a special bipartite entanglement spectrum of atomic quantum Hall states, and
enables an intrinsic probe of these highly correlated states using
two-body short-ranged correlations.

\end{abstract}

\maketitle

Ultracold atoms 
interact with short-range interactions $U({r})$, which vanishes when the separation between two atoms is larger than a length scale $r_0$. The atomic density in typical experiments is very low such that $k_F\ll r_0^{-1}$ is well satisfied. 
As originally discovered by Shina Tan\cite{Tan1, Tan2, Tan3}, such diluteness leads to a wide range of fundamental relations among thermodynamic quantities in the presence of $s$-wave scattering. These universal relations are governed by contact $C$. 
The momentum distribution $n({\bf k})$ has an asymptotic behavior $C/|{\bf k}|^4$ at large $|{\bf k}|$. The same $C$ also shows up in the energy functional, the adiabatic relation, dynamic structure factors, and other relations \cite{V1,V2,V3,V4}. Many of these universal relations have been verified in experiments, and contact has been continuously inspiring physicists in both the ultracold atom and nuclear physics community to explore its applications\cite{Jin1, Jin2, Jin3,Vale,T1,T2,T3,T4,T5, Zhou,Drut}.

In this Letter, we show that the concept of contact can be generalized to an arbitrary partial-wave scattering, and we define a contact spectrum $\{C_{lm}\}$ formed by contact in all partial-wave channels, where $(l, m)$ are the quantum numbers for the angular momentum. $\{C_{lm}\}$  controls universal relations, such as the large momentum distribution and the energy functional,  in the presence of arbitrary partial-wave scatterings, which have considerable differences from those in the presence of $s$-wave scattering alone, due to the fundamental differences between high-partial-wave scatterings and the $s$-wave one. In addition to thermodynamic relations,  $\{C_{lm}\}$ provides physicists a new means to trace many-body physics from short-range quantities in strongly correlated systems.

\begin{figure*}
\includegraphics[width=7in]{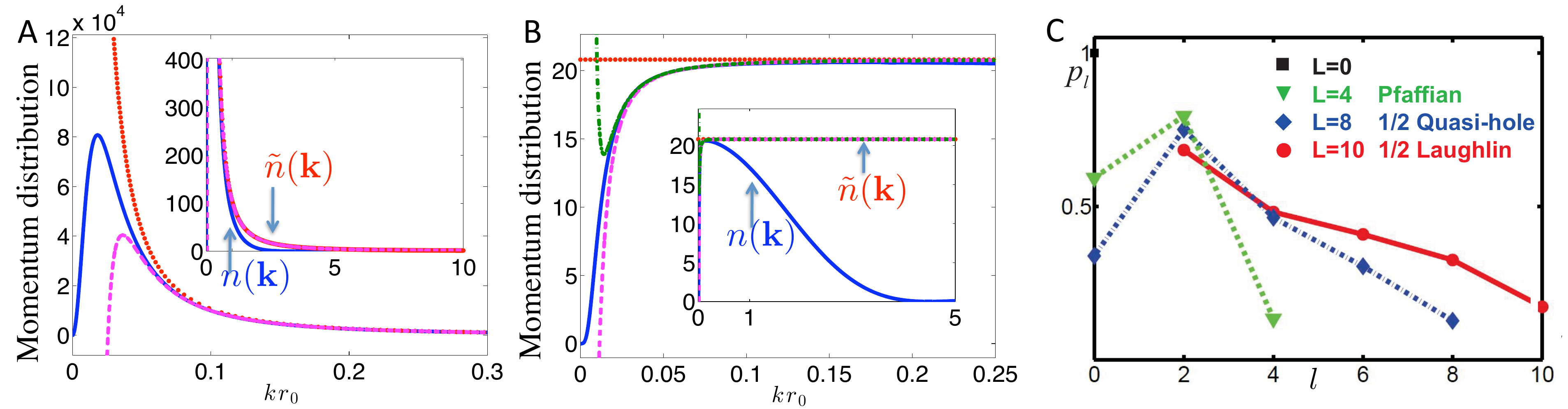}
\caption{(A-B), momentum distributions in unit of $r_0^3$ of $p$- and $d$-wave  two-body bound states respectively. Solid blue curves are exact results, dotted red ones represent the leading term $k^{2l-4}$ alone. Dash (pink) and dash dotted (green) curves include contributions up to the subleading and the third leading terms respectively. Insets compare the realistic momentum distribution $n({\bf k})$ and $\tilde{n}({\bf k})$ obtained from extending the universal behavior of $n({\bf k})$ in the regime  $k\ll r_0^{-1}$ to infinity in the zero-range interaction approximation limit. (C) $\{p_l\}$ for different bosonic QAH states of $4$ atoms. With increasing the total angular momentum $L$, $\{p_l\}$ becomes broader and $p_{l<m}$ eventually vanish in the Laughlin state with filling factor $1/2$. }
\end{figure*}

We consider the many-body wave function of a single-component atoms, either bosons or fermions, with a fixed total angular momentum $L$, $\Psi_{3D}\equiv \Psi({\bf r}_1, {\bf r}_2, ..., {\bf r}_N)$, where $N$ is the total particle number. Due to the length scale separation $r_0\ll k_F^{-1}$, $\Psi_{3D}$ has a unique asymptotic behavior when the distance between two particles $|{\bf r}_{ij}|$ is much smaller than $k_F^{-1}$,
 \begin{equation}
\Psi_{3D} \stackrel{|{\bf r}_{ij}|\ll k_F^{-1}}{\xrightarrow{\hspace*{1cm}} } \sum_{lm} \int d\epsilon  \psi_{lm}({\bf r}_{ij};\epsilon)G_{lm}({\bf R}_{ij};E-\epsilon) \label{Asym}
\end{equation}
where ${\bf r}_{ij}={\bf r}_i-{\bf r}_j$, ${\bf R}_{ij}=\{\frac{{\bf r}_i+{\bf r}_j}{2},{\bf r}_{k\neq i,j}\}$,  $E$ is the total energy, $\psi_{lm}({\bf r}_i-{\bf r}_j;\epsilon)$ is an unnormalized solution, which satisfies the boundary condition at ${\bf r}_i={\bf r}_j$, for the relative motion of the two particles with angular momentum quantum number $(l,m)$ and energy $\epsilon$ under the two-body Hamiltonian, $ H^{[2]}=-\frac{\hbar^2}{M}\nabla^2+U(|{\bf r}_{ij}|)$.  The symmetric or antisymmetric nature of the bosonic or fermionic wave function automatically picks up even or odd $l$ in the summation.  Equation (\ref{Asym}) can be understood from that other atoms  essentially cannot affect the relative motion of the $i$th and $j$th atoms when ${|{\bf r}_{ij}|\ll k_F^{-1}}$. $G_{lm}({\bf R}_{ij};E-\epsilon) $ characterizes the center of the mass of this pair of atoms and the other $N-2$ ones, and acts as a ``normalization factor" for the pair wave function  $\psi_{lm}({\bf r}_i-{\bf r}_j;\epsilon)$. 
Unlike an isolated two-body system, neither the angular momentum $(l, m)$ nor the energy $\epsilon$ is conserved for any pair of particles in many-body systems.

 Whereas $\psi_{lm}({\bf r}_{ij};\epsilon)$ in the region $r<r_0$ depends on the microscopic details of the potential $U(r)$, it takes a simple and universal form for $r>r_0$ where $U(r)=0$,
\begin{equation}
\psi_{lm}({\bf r}_{ij};\epsilon)\sim\{j_l(q_\epsilon r_{ij})-\tan[\eta_l(q_\epsilon)]n_l(q_\epsilon r_{ij})\}Y_{lm}(\hat{\bf r}_{ij})
\end{equation}
where $q_\epsilon^2/M=\epsilon$, $j_l(x)$ and $n_l(x)$ are spherical Bessel functions of the first and second kinds respectively, $Y_{lm}(\hat{\bf r}_{ij})$ is the spherical harmonics and $\hat{\bf r}_{ij}$ is a unit vector.  Using asymptotic  forms of $j_l(x) $ and $n_l(x)$ at small $x$,  we see that at small $|{\bf r}_{ij}|$,
\begin{equation}
\begin{split}
\psi_{lm}({\bf r}_{ij};\epsilon)\rightarrow &\Big[q_\epsilon^{2l+1}{\cot[\eta_l(q_\epsilon)]}\sum_{s=0}\alpha_{ls} q_\epsilon^{2s} r_{ij}^{l+2s}\\&+\sum_{s=0}\frac{\beta_{ls} q_\epsilon^{2s} }{(r_{ij})^{l+1-2s}}\Big]Y_{lm}(\hat{\bf r}_{ij}), \label {S2}
\end{split}
\end{equation}
where $\alpha_{ls}=\frac{(-1)^{s}}{(2s)!!(2l+2s+1)!!}$, $\beta_{ls}=\frac{(2l-2s-1)!!}{(2s)!!}$.  The phase shift has a standard low energy expansion,  $q^{2l+1}_\epsilon{\cot[\eta_l(q_\epsilon)]}=(-{a_l}^{-1}+r^e_l q^2_\epsilon+...)r_0^{-2l}$, where $a_l$ and $r^e_l$ are the scattering length and the effective range of the $l$th partial-wave\cite{Baym}. 
For $s$-wave scattering in broad resonances, $r^e_0$ is not important, and $\psi_{00}({\bf r}_{ij};\epsilon)$ has the conventional asymptotic form $(\frac{1}{r_{ij}}-\frac{1}{a_0})$ at small distance.

An angular-momentum-selective contact is defined for a given $l$-th partial-wave as\begin{equation}
{C}_{lm}=(4\pi)^2 N(N-1)\int d {\bf R}_{ij}\Big|\int d\epsilon G_{lm}({\bf R}_{ij};E-\epsilon)\Big|^2,
\end{equation}
where $\int d {\bf R}_{ij}= \int d(\frac{{\bf r}_{i}+{\bf r}_j}{2})d{\bf r}_{k\neq i, j}$. Using
\begin{equation}
n_{\bf k}=\sum_{i}\int \prod_{j\neq i}d{\bf r}_{j}\Big|\int d{\bf r}_i\Psi({\bf r}_1,{\bf r}_2,... {\bf r}_N)e^{i {{\bf k}\cdot {\bf r}_i}}\Big|^2,
\end{equation}
a straightforward calculation as presented in the Supplementary Materials shows that
\begin{equation}
n_{\bf k}=\sum_{lm} n_{lm}(k)|Y_{lm}(\hat{\bf k})|^2,\label{nk}
\end{equation}
where the asymptotic form of $ n_{lm}(k)$ is written as
\begin{equation}
n_{lm}(k)\stackrel{k_F\ll k \ll r_0^{-1}}{\xrightarrow{\hspace*{1cm}} } C_{lm}k^{2l-4}+ \sum^l_{s=1} {C}^s_{lm}k^{2l-4-2s} \label{nks}.
\end{equation}
In the presence of $s$-wave scattering alone,  the expression reduces to the well known result $n_{\bf k}\rightarrow C_{00}k^{-4} |Y_{00}(\hat{\bf k})|^2$, where $|Y_{00}(\hat{\bf k})|^2=1/(4\pi)$ leads to a trivial pre-factor difference of $C_{00}$ from the original $s$-wave contact defined by Tan. For  any given high partial-wave scattering $l>0$, the leading term in the large momentum distribution $C_{lm}k^{2l-4}$ is distinct from the $s$-wave one. It also contains the contributions from the subleading and other terms, $C^1_{lm}k^{2l-6}$ and etc.  Expressions for $C_{lm}^{s}$ are given in Supplementary Materials.  Equation (\ref{nks}) can be demonstrated using a two-body problem, as shown in figure (1A-1B) for a simple square well interaction potential.

Equation (\ref{nks}) tells one  that, if a zero range potential approximation $r_0\rightarrow 0$ is applied, an ultraviolet divergence of energy shall occur. For instance, the leading term $k^{2l-4}$ in the $l$th partial-wave alone leads to $\int^{\Lambda}d{\bf k} k^2 n_{lm}(k) Y_{lm}(\hat{\bf k})\sim \Lambda^{2l+1}$, where $\Lambda$ is the large momentum cutoff. For $l=0$, Shina Tan has invented a remarkable functional integral form  to remove such a divergence, which reveals the universal relation between the internal energy of dilute many-body systems and $s$-wave contact\cite{Tan1,Tan2, Tan3}. Despite that the scattering in a high-partial-wave has fundamental differences with the $s$-wave one,  a similar energy functional could be derived in the presence of arbitrary partial-wave scatterings, \begin{equation}
\begin{split}
& E P_{r_0}=\int \frac{d{\bf k}}{(2\pi)^3}\frac{\hbar^2k^2}{2M}\Big[\tilde{n}({\bf k})-\sum_{lm}\left(\frac{C_{lm}}{k^{4-2l}}+\sum_{s=1}^l\frac{ C^s_{lm}f_l^s(k r_0)}{k^{4-2(l-s)}} \right)\\ &{ {|Y_{lm}(\hat{\bf k})|^2}}\Big]  +\sum_{lm}\Big(u_l \frac{C_{lm}}{a_l}+v_l r^e_l C^1_{lm}\Big)+O((k_F r_0)^2) \label{efu},
\end{split}
\end{equation}
where $E$ is the internal energy, $\tilde{n}({\bf k})$ is the momentum distribution by extending its universal behaviors in the regime $k_F\ll k\ll 1/r_0$ to infinity, as shown in figure (1A-1B), and $u_l=\frac{\hbar^2 r_0^{-2l}}{32\pi^2 M}$, $v_l=-\frac{\hbar^2 r_0^{-2l}}{64\pi^2 M}$, and $f_l^s(k r_0)$ are a set of functions that depend on the range of interaction $r_0$. The details of the derivation of the energy functional, as well as the explicit expressions of  $f_l^s(k r_0)$, are given in the Supplementary Materials. In the regime $kr_0\gg 1$, for any $s$, $f_l^s(k r_0)\rightarrow 1$. Using equations (\ref{nk}) and (\ref{nks}), one sees that the divergence in the kinetic energy encoded in $\tilde{n}({\bf k})$ is removed.  On the left hand side of the equation (\ref{efu}), a factor $P_{r_0}=1-N(N-1) I_{r_0}/2$ is included, where
\begin{equation}
I_{r_0}=\sum_{lm}\int d{\bf R}_{ij} \int_0^{r_0} d{\bf r}_{ij} |\int d\epsilon\psi_{lm}({\bf r}_{ij};\epsilon) G_{lm}({\bf R}_{ij}, E-\epsilon)|^2
\end{equation}
characterizes the total weight of the two-body wave function in the volume defined by $|{\bf r}_{ij}|<r_0$. It can also be expressed using $C_{lm}$ (Supplementary Materials). We have also verified this energy functional using the two-body problem which can be solved exactly.


Whereas equation (\ref{efu}) is similar to the one for the $s$-wave scattering, it has a few notable features. First, the effective range $r^e_l$ explicitly enters the expression, since it is required for describing microscopic physics of a high partial-wave scattering, such as the bound state energy $\sim 1/(a_l r_l^{e})$. 
Second, the terms ${C^{s}_{lm}}$ must be included for completely removing the divergence for $l>0$, where all terms in equation (\ref{nks}) contribute to the divergence.  Finally, the singular part of the two-body wave function $\sim {\bf r}_{ij}^{-l-1}$ in equation (\ref{S2}) is not normalizable for $l>0$ in the zero range potential approximation that takes $r_0\rightarrow 0$. 
$P_{r_0}$ is therefore required in equation (\ref{efu}). In the presence of $s$-wave scattering alone, the zero range approximation can be safely applied and $P_{r_0}$ is negligible.


All the above discussions can be directly generalized to two dimensions. One simply needs to replace $\psi_{lm}({\bf r}_{ij};\epsilon)$ in equation (\ref{Asym}) by the solution of the two-body Hamiltonian in two dimensions. For $|z_{ij}|>r_0$,  where $z_i=x_i+i y_i$ is the coordinate of the $i$th atom in two dimensions, such solutions are simply Bessel functions, i.e., $\sim [\cot \eta_l(q_\epsilon) J(q_\epsilon |z_{ij}|)- Y_l(q_\epsilon |z_{ij}|)]e^{il\theta_{ij}}$, where $\eta_l(q_\epsilon) $ is the two-dimensional phase shift, and $\theta_{ij}=\arg\{z_{ij}\}$. In the regime ${r_0<|z_{ij}|\ll k_F^{-1}}$, $\Psi_{2D}\rightarrow A_0({\bf Z}_{ij})\ln(\frac{|z_{ij}|}{a_0})+\Psi_{2D}^{l\neq 0}$,
\begin{equation}
\Psi_{2D}^{l\neq 0}=\sum_{l\neq 0}  \left(A_l({\bf Z_{ij}})|z_{ij}|^l+\frac{B_l({\bf Z_{ij}})}{|z_{ij}|^l}\right)e^{il\theta_{ij}},\label{BC2D}
\end{equation}
where $
A_l({\bf Z_{ij}})=-\frac{1}{a_l}\frac{r^{-2l+1}_0}{(2l)!!}\int d\epsilon {{G_{l}}}({\bf Z}_{ij};E-\epsilon)$ and $B_l({\bf Z_{ij}})=\frac{2}{\pi} (2l-2)!! \int d\epsilon {{G_{l}}}({\bf Z}_{ij};E-\epsilon)$, so that
\begin{equation}
A_l({\bf Z_{ij}})=B_l({\bf Z_{ij}}) a_l^{-1}\frac{-r^{-2l+1}_0\pi}{2(2l)!!(2l-2)!!}, \label{AB}
\end{equation}
where ${\bf Z}_{ij}=\{\frac{z_i+z_j}{2}, z_{k\neq i, j}\}$.

To simplify notations, we keep only the leading terms in the expansions of the Bessel functions. 
Similar to three dimensions, we define contact in an arbitrary high partial-wave channel,
\begin{equation}
C_{l\neq 0}=\frac{(2\pi)^2}{\left[(2l-2)!!\right]^2} N(N-1)\int d{\bf Z}_{ij}|B_l({\bf Z_{ij}})|^2,\label{C2D}
\end{equation}
where $\int d{\bf Z}_{ij}=\int dZ_{ij}\prod_{k\neq ij}dz_k$. The leading term in the large momentum distribution is given by $C_{l}k^{2l-4}$, which has the same power as that in three dimensions. Whereas the energy functional can be also derived in the same manner as that for three dimensions, we here focus on the application of contact spectrum in two dimensions in atomic quantum Hall states (QHS).





QHS are intriguing many-body states carrying a large amount of angular momenta.  At filling factor of $1/m$, QHS  is characterized by the  Laughlin wave function\cite{Laughlin},
\begin{equation}
\Psi_L= \prod_{i>j}(z_i-z_j)^me^{-\sum_i|z_i|^2/4\sigma^2} / \sqrt{\mathcal{N}} ,\label{LS}
\end{equation}
where $\mathcal{N}$ is the normalization factor, $m$ is odd(even) for fermions(bosons),  $\sigma$ is the magnetic length. $\Psi_L$ carries a total angular momentum $L=m N(N-1)/2$. 
There are two steps for establishing the applications of contact spectrum in atomic QHS.

{\bf Step 1}  We rewrite $\Psi_L$ as
\begin{equation}
\Psi_L=\sum_{l=m}^{m(2N-3)}p_l\left(\lambda_l^{-1}z_{ij}^le^{-\frac{|z_{ij}|^2}{4\sigma^2}}\right)F_{L-l}({\bf Z}_{ij}), \label{LSA}
\end{equation}
where  $z_{ij}=|z_{i}-z_{j}|e^{i\theta_{ij}}$, $\lambda_l=\sigma^{l+1}\sqrt{2^{l+1}\pi l!}$ so that the two-body wave function in the parentheses is normalized to 1,  and $F_{S}({\bf Z}_{ij}) $ is a set of orthogonal normalized wave function with angular momentum $S$ and depends on $(z_i+z_j)/2$ and the other $N-2$ particles. 

{\bf Step 2} Though $\Psi_L$  is the exact ground state if one sets the interaction in all partial-wave channels with $l\ge m$ to be zero in the framework of Haldane pseudopotential\cite{Haldane}, it is in general  a trial wave function for realistic systems. The exact ground state with the same total angular momentum $\Psi^{ex}_L$ must contain corrections to $\Psi_L$.  For electrons with long-range Coulomb interaction, a standard approach is to numerically compute the overlap integral  $\langle \Psi_L|\Psi_L^{ex}\rangle$. In dilute systems, it requires that $\Psi_L^{ex}$ satisfies the boundary condition at short distance in equation (\ref{BC2D}) and recovers $\Psi_L$ at large distance $|{ z}_{ij}|$.

Equation (\ref{LSA}) shows that each Laughlin state has a unique distribution of $\{p_l\}$,  with finite $p_l$  between $l_{min}=m$ and $l_{max}=m(2N-3)$. The very broad distribution of $\{p_l\}$ as reflected by  $l_{max}-l_{min}\sim N$ is a direct consequence of strong many-body entanglement in QHS. 
The same strategy can be applied to other types of QHS, such as the Pfaffian and quasi-hole states, described by $\Psi_{pf}={Pf\left(\frac{1}{z_i-z_j}\right)\prod_{i>j}(z_i-z_j)e^{-\sum_i|z_i|^2/4\sigma^2}}$ and $\Psi_h={\frac{\partial}{\partial z_1}\frac{\partial}{\partial z_2}\cdots\frac{\partial}{\partial z_N}\left(\prod_{i>j}(z_i-z_j)^m\right)e^{-\sum_i|z_i|^2/4\sigma^2}}$ respectively, where $Pf\left(\frac{1}{z_i-z_j}\right)$ is Pfaffian of $(z_i-z_j)^{-1}$. Figure(1C) shows  $\{p_l\}$ for a few bosonic QAH states.

Equation (\ref{LSA}) also allows one to define a special bipartite entanglement spectrum. For a pure state wave function, $|\Psi\rangle=\sum_\lambda (e^{-\xi_\lambda/2}/\sqrt{Z}) |\Psi^{R}_\lambda\rangle|\Psi^{L}_\lambda\rangle$,  where $Z=\sum e^{-\xi_\lambda}$, and $|\Psi^{R}_\lambda\rangle$ and $|\Psi^{L}_\lambda\rangle$ are the eigenstates of its two subsystems, $\{\xi_\lambda\}$ defines the entanglement spectrum\cite{ES}.  Using  equation (\ref{LSA}), one observes that $\{\xi_l\equiv-2\log p_l-\log Z\}$ characterizes the entanglement spectrum between the relative motion of a pair of atoms and the rest of the system in the angular momentum space. Whereas  entanglement is usually considered by spatially dividing the system into two parts,  $\{p_l\}$ here provides one a new angle to characterize QHS, 
since 
$\{p_l\}$ captures the correlation in the angular momentum space, 
regardless of the distance $|z_{ij}|$ between particles. Thus, the signature of many-body entanglement persists in short distance and is captured by the contact spectrum as discussed below.

When $|z_{ij}|$ approaches zero, equation (\ref{LSA}) becomes
\begin{equation}
\Psi_L\stackrel{ |z_{ij}|\ll \sigma}{\xrightarrow{\hspace*{0.8cm}} }  \sum_{l=m}^{m(2N-3)}  \lambda_l^{-1}p_lF_{L-l}({\bf Z}_{ij})|z_{ij}|^le^{il\theta_{ij}}\label{LSB},
\end{equation}
which does not satisfy the boundary condition specified by equation (\ref{BC2D}). 
Though the high partial-wave scattering length in general is smaller than the $s$-wave one, for realistic interaction with a finite range $r_0$, the singular terms $1/|z_{ij}|^l$ must exist and becomes important in short distance. In ultracold atoms, $a_{l\neq 0}$ can be tuned by Feshbach resonance and other techniques, so that the singular terms could have observable effects even away from the resonance.


Whereas $\langle \Psi_L|\Psi_L^{ex}\rangle$ can also be computed for short-range interactions, the diluteness provides one a simple method to qualitatively estimate such an overlap integral. 
We define a length scale $r^*=\max \{r_l^*\}$, where $r_l^*$ satisfies $|A_l| r_l^{*l}=|B_l|{r_l^{*-l}}$. It is easy to verify that $r_l^*\sim r_0(a_l/r_0)^{\frac{1}{2l}}$ for low energy scattering $q_{\epsilon}\ll r_0^{-1}$. If the separation between any paired atoms is much larger than $r^*$, the correction to the  wave function due to the weak scattering $a_l\ll \sigma$ in all partial-wave channels becomes negligible. Using equation (\ref{BC2D}), the exact wave function  becomes
\begin{equation}
\Psi_L^{ex}\stackrel {r^*\ll |z_{ij}|\ll \sigma}{\xrightarrow{\hspace*{1cm}} } \sum_{l\neq 0} {A_l}({\bf Z}_{ij}) |z_{ij}|^le^{il\theta_{ij}},\label{ES}
\end{equation}

Compare equation (\ref{LSB}, \ref{ES}), one see that,  if the following equation
\begin{equation}
A_l({\bf Z}_{ij})= \lambda_l^{-1}p_lF_{L-l}({\bf Z}_{ij}),\label{Al}
\end{equation}
is satisfied,  the exact ground state wave function $\Psi_L^{ex}$ reduces to $\Psi_L$ at large inter-particle distance, $\Psi_L^{ex}\stackrel{ |z_{ij}|\gg r^*}{\rightarrow} \Psi_L$. The overlap between $\Psi_L^{ex}$ and $\Psi_L$ can be estimated as $1-O(k_F^2r^{*2})$, where $k_F\sigma\sim 1$ in QHS.  The criterion for $\Psi_L$ to be a good approximation of the exact ground state is that $r^*\ll k_F^{-1}$. 
If the scattering for all $l\ge m$ vanish, $r_{l\ge m}^*=0$, and $\Psi_L$ becomes the exact ground state, consistent with results from Haldane pseudopotential.

Using equations (\ref{AB},\ref{C2D},\ref{Al}), one establishes an intrinsic relation between contact spectrum $\{C_{l}\}$ and the entanglement spectrum $\{p_l\}$,
\begin{equation}
p_l= \frac{r^{-2l+1}_0}{2^{l+2} l!}\frac{a_l^{-1} \lambda_l}{\sqrt{N(N-1)}}C_{l}^{1/2} \label{td}.
\end{equation}
As contact spectrum can be measured through various schemes,  equation (\ref{td}) allows one to probe $\{p_l\}$. 
Besides the large momentum distribution measured in Time-Of-Flight experiments, another useful scheme for measuring $\{C_{l}\}$ is photoassociation, which can be made angular-momentum-selective. For instance, by tuning the laser to be resonant with an excited state with a particular angular momentum $l_e\neq0$, only the  $l=l_e$th partial-wave of the relative motion of two atoms contributes to the photoassociation.  For a two-body problem, an important quantity to characterize the stimulated rate of populating the excited state and the scattering lengths of the optical Feshbach resonance is the coupling strength to the excited state, 
$\Gamma^{[2]}_l=c^{-1}({2\pi I})d_M^2|\langle\varphi_{e,l} |\phi^{[2]}_l \rangle|^2$\cite{PA1, PA2, PA3}, where 
$c$ is the speed of light, $I$ is laser intensity,  $d_M$ is the dipole moment, $\varphi_{e,l}$ 
is the electronically excited molecular wave function 
with angular momentum $l$,  and $\phi^{[2]}_l$ is the relative wave function of two atoms. 

In dilute many-body systems, the size of the electronically excited molecules is much smaller than inter particle spacing. 
Since $G_{l}({\bf Z}_{ij};E-\epsilon)$ (or $G_{lm}({\bf R}_{ij};E-\epsilon)$ in three dimensions) acts as the normalization factor for the two-body wave function $\psi_l(z_{ij},\epsilon)$, the coupling strength in the many-body system can be written as
\begin{equation}
\Gamma_l= \frac{\pi I}{16 c}d_M^2 {C}_{l}|\langle\varphi_{e,l}| \psi_{l}^0  \rangle|^2,
\end{equation}
where a low-energy expansion $\psi_{l}(z_{ij},\epsilon )=\sum_s \epsilon^{s}\psi^s_{l}(z_{ij})$ has been used. In an experiment by Gemelke, et al, photoassociation has been used for detecting atomic QHS clusters\cite{Nate}. Whereas the decreased signal of photoassociation with increasing the rotation frequency is consistent with accessing the QAH regimes, an angular momentum selective photoassociation can further probe the rise of high partial-wave contact, so that much richer information of atomic QAH states can be obtained, as shown by figure (1C) and equation (\ref{td}).

In conclusion, we have generalized the concept of contact to arbitrary partial-waves, and defined contact spectrum $\{C_{lm}\}$ and $\{C_{l}\}$ in 3D and 2D. In addition to thermodynamic relations, we have applied $\{C_{l}\}$ in atomic QHS, and shown that it serves as an intrinsic probe of atomic QHS due to a connection with a special entanglement spectrum of QHS. We hope that this work may stimulate more studies on the connection between short-range correlations and many-body physics in strongly correlated dilute systems.

Note: near the completion of this manuscript,   two theoretical and one experimental paper \cite{P1, P2, P3}  addressing contact of a single high partial-wave scattering ($l=1$ case in our work) showed on arxiv. All works agree on the leading term in the large momentum distribution for the p-wave scattering, and \cite{P2, P3} have also explored the subleading term  consistent with ours. QZ acknowledges D. Wang, J. H. Thywissen, S. Zhang and J. Zhang for discussions. This work is supported by RGC/GRF(14306714).


*MH and SZ contribute equally to this work.

\onecolumngrid

\vspace{0.4in}

\centerline{\bf Supplementary Material}

\vspace{0.2in}

In this supplementary material, we present the results on the large momentum distribution and the energy functional.

\vspace{0.15in}

{\bf Large momentum distribution}

Starting from $n_{\bf k}=\sum_{i}\int \prod_{j\neq i}d{\bf r}_{j}\Big|\int d{\bf r}_i\Psi({\bf r}_1,{\bf r}_2,... {\bf r}_N)e^{i {{\bf k}\cdot {\bf r}_i}}\Big|^2$,
we define $\psi_i({\bf k})=\int d {\bf r}_i\Psi({\bf r}_1,{\bf r}_2,... {\bf r}_N)e^{i {{\bf k}\cdot {\bf r}_i}}$ so that $n_{\bf k}=\sum_{i}\int \prod_{j\neq i}d{\bf r}_{j}|\psi_i({\bf k})|^2$ . 
We also define a purely two-body quantity,
\begin{equation}
F_{lm}({\bf k};\epsilon)=\left(\int d{\bf r}_{ij}{\psi}_{lm}({\bf r}_{ij},\epsilon)e^{i {{\bf k}\cdot ({\bf r}_i-{\bf r}_j)}}\right)
\end{equation}
so that in the regime $|{\bf k}|\gg k_F$ can be written as $\psi_i({\bf k})\rightarrow \sum_{j\neq i}e^{i {{\bf k}\cdot {\bf r}_j}}\sum_{lm}\int d\epsilon G_{lm}({\bf R}_{ij};E-\epsilon)F_{lm}({\bf k};\epsilon)$.
If one extends the wave function ${\psi}_{lm}({\bf r}_{ij},\epsilon)$ in the regime $[r_0,\infty] $ to $[0, \infty]$,  $F({\bf k};\epsilon)$  in the regime $k_F\ll k$  is given by,
\begin{equation}
F_{lm}({\bf k};\epsilon)=\left(\int_{0}^\infty d{\bf r}_{ij} \left( \frac{\beta_{l0}}{r_{ij}^{l+1}} +\frac{\beta_{l1} q^2_\epsilon}{r_{ij}^{l-1}}+...\right)Y_{lm}(\hat{\bf r}_{ij})e^{i {{\bf k}\cdot ({\bf r}_i-{\bf r}_j)}}\right)
\end{equation}
For s-wave scattering, we have
\begin{equation}
F_{00}({\bf k};\epsilon)={4\pi}/k^2
\end{equation}

For $l>0$, using  $e^{i{\bf k}\cdot{\bf r}}=4\pi\sum_{l=0}^{\infty}\sum_{m=-l}^{l}i^lj_l(kr)Y_{lm}({\bf k}/k)Y^*_{lm}({\bf r}/r)$,  we  obtain
\begin{eqnarray}
\psi_i({\bf k})\rightarrow \sum_{j\neq i}e^{i {{\bf k}\cdot {\bf r}_j}}\sum_{lm}4\pi i^l\Big(k^{l-2}\int d\epsilon G_{lm}({\bf R}_{ij};E-\epsilon) + k^{l-4}\int d\epsilon q^2_\epsilon G_{lm}({\bf R}_{ij};E-\epsilon) +...\Big)Y_{lm}(\hat{\bf k}) ,
\end{eqnarray}
When calculating the integral $\int \prod_{j\neq i}d{\bf r}_j|\psi_i({\bf k})|^2$, the cross term $e^{i{{\bf k} \cdot (\bf r_{j'}-{\bf r}_{j})}}$ vanishes in the large $k$ limit. Changing $d{\bf r}_{j\neq i}$ to $d{\bf R}_{ij}$, the cross term $G^*_{l' m'}({\bf R}_{ij},E-\epsilon_q)G_{l m}({\bf R}_{ij},E-\epsilon_{q'})$ also vanishes, due to the orthogonality of wave functions with different angular momenta,  we obtain
\begin{equation}
\tilde{n}_{\bf k}=k^{-4}C_{00} |Y_{00}(\hat{\bf k})|^2+ \sum_{l \neq 0, m}  (k^{2l-4}C_{lm} +  k^{2l-6}C^{1}_{lm}+ k^{2l-8}C^{2}_{lm}+...) |Y_{lm}(\hat{\bf k})|^2,
\end{equation}
$C_{lm}$ has been defined in the  main text, and
\begin{equation}
C^{1}_{lm}=32\pi^2 {N(N-1)}  Re \Big[ \int d{\bf R}_{ij}\left(\int d\epsilon G^*_{lm}({\bf R}_{ij};E-\epsilon) \right)\left(\int d\epsilon q^2_\epsilon G_{lm}({\bf R}_{ij};E-\epsilon)\right) \Big]
\end{equation}
\begin{equation}
C^{2}_{lm}=32\pi^2 {N(N-1)} \int d{\bf R}_{ij}\Big\{  Re \Big[\left(\int d\epsilon G^*_{lm}({\bf R}_{ij};E-\epsilon) \right)\left(\int d\epsilon q^4_\epsilon G_{lm}({\bf R}_{ij};E-\epsilon)\right) \Big]+\frac{1}{2}\Big|\int d\epsilon q^2_\epsilon G_{lm}({\bf R}_{ij};E-\epsilon)\Big|^2\Big\}.
\end{equation}
Other $C_{lm}^{s}$ can also be written down straightforwardly using the same procedures. \\

{\bf Energy functional}

In the regimes where the distance between any pair of particles is larger than $r_0$, the many-body Schr\"odinger equation becomes $
\left(\sum_i -\frac{\hbar^2}{2M}\nabla_i^2\right)\Psi =E\Psi$, where the arguments $({\bf r}_1, {\bf r}_2, ..., {\bf r}_N)$ has been suppressed. It leads to
\begin{equation}
\int' \prod_id{\bf r}_i\Psi^*\left(\sum_i -\frac{\hbar^2}{2M}\nabla_i^2\right)\Psi =E\int' \prod_id{\bf r}_i | \Psi |^2\label{En}
\end{equation}
where $\int' \prod_id{\bf r}_i$ means the integration is carried out in the regions where any pair of particles is larger than $r_0$.

In the zero range interaction limit, one extends the asymptotic form
\begin{equation}
\Psi\rightarrow \sum_{lm}\int d\epsilon \left(\frac{q_\epsilon^{l+1}}{\tan[\eta_l(q_\epsilon)]}j_l(q_\epsilon r_{ij})-q_\epsilon^{l+1}n_l(q_\epsilon r_{ij})\right)G_{lm}({\bf R}_{ij};\epsilon)Y_{lm}(\hat{\bf r}_{ij}),
\end{equation}
which is valid in the regime $r_0\ll r_{ij}<k_F^{-1}$ to the regime $r_{ij}<r_0$.  Define $\Phi$, which is identical to $\Psi$ if $r_{ij}>r_0$ and given by the right hand side (RHS) of the above equation,  the integral on the left hand side (LHS) of equation (\ref{En}) can be extended to the whole real space, and meanwhile, the integral in the unphysical region $|r_{ij}|< r_0$ must be subtracted to remove the divergence. If three-body physics is ignored, we obtain,
\begin{equation}
\begin{split}
LHS&=\int \prod_id{\bf r}_i\Phi^*\left(\sum_i -\frac{\hbar^2}{2M}\nabla_i^2\right)\Phi-\frac{N(N-1)}{2}\int \prod_{k\neq i,j}  d{\bf r}_k\int d\left(\frac{{\bf r}_{i}+{\bf r}_i}{2}\right)\int_0^{r_0} d{\bf r}_{ij}\Phi^* \left(-\frac{\hbar^2}{M}\nabla_{ij}^2\right)\Phi \\
&=\int d{\bf k} \frac{\hbar^2k^2}{2M} \tilde{n}_{\bf k}
-\frac{N(N-1)}{2}\int d{\bf R}_{ij}\int_0^{r_0} d{\bf r}_{ij}\Phi^* \left(-\frac{\hbar^2}{M}\nabla_{ij}^2\right)\Phi\\
&=\int d{\bf k} \frac{\hbar^2k^2}{2M} \tilde{n}_{\bf k}+\frac{N(N-1)}{2}\frac{\hbar^2}{M}\sum_{lm}\int d{\bf R}_{ij}  \int_0^{r_0}d{\bf r}_{ij}\Big\{\Big[Y^*_{lm}(\hat{\bf r}_{ij}) \int d\epsilon G^*_{lm}({\bf R}_{ij};\epsilon) \left(\frac{q_\epsilon^{l+1}}{\tan[\eta_l(q_\epsilon)]}j_l(q_\epsilon r_{ij})-q_\epsilon^{l+1}n_l(q_\epsilon r_{ij})\right)^\ast \Big]  \\
&\nabla_{ij}^2 \Big[ \int d\epsilon G_{lm}({\bf R}_{ij};\epsilon) \left(\frac{q_\epsilon^{l+1}}{\tan[\eta_l(q_\epsilon)]}j_l(q_\epsilon r_{ij})-q_\epsilon^{l+1} n_l(q_\epsilon r_{ij})\right)Y_{lm}(\hat{\bf r}_{ij})\Big]\Big\}\label{rd}
\end{split}
\end{equation}

As seen from the asymptotic form of $\tilde{n}({\bf k})$ at large $|{\bf k}|$, we conclude that the integral $\int d{\bf k} \frac{\hbar^2k^2}{2M} \tilde{n}_{\bf k}$ is divergent. It is worth mentioning that not only the leading term $k^{2l-4}$ contributes to the divergence, many other terms for large $l$ will also contributes to the divergence. The divergence therefore needs to be carefully removed.

One term in the last line of equation (\ref{rd}) automatically removes the divergence.  To see this fact, we examine the contributions separately from $j_l(q_\epsilon r_{ij})$ and $n_l(q_\epsilon r_{ij})$.
Since $\nabla_{ij}^2 j_l(q_\epsilon r_{ij}) Y_{lm}(\hat{\bf r}_{ij})=-q_\epsilon^2 j_l(q_\epsilon r_{ij})Y_{lm}(\hat{\bf r}_{ij})$, one sees that
\begin{equation}
 \int_0^{r_0}d{\bf r}_{ij}\Big\{\Big[Y^*_{lm}(\hat{\bf r}_{ij})  \left(\frac{q_\epsilon^{l+1}}{\tan[\eta_l(q_\epsilon)]}j_l(q_\epsilon r_{ij})-q_\epsilon^{l+1}n_l(q_\epsilon r_{ij})\right)^\ast\Big] \nabla_{ij}^2 \frac{q_\epsilon^{l+1}}{\tan[\eta_l(q_\epsilon)]} j_l(q_\epsilon r_{ij}) Y_{lm}(\hat{\bf r}_{ij})\sim \frac{q_\epsilon^{2l+1}}{\tan[\eta_l(q_\epsilon)]}(q_\epsilon r_0)^{2}
\end{equation}
In the low energy limit, $(q_\epsilon r_0)^{2}\ll 1$, such a contribution is negligible.

In contrast, $\nabla_{ij}^2 n_l(q_\epsilon r_{ij})Y_{lm}(\hat{\bf r}_{ij})$ is crucial. By making use of $\left(\frac{1}{r^2}\frac{d}{d r}r^2\frac{d}{dr}-\frac{l(l+1)}{r^2}\right) \frac{1}{r^{l+1}}=-(2l+1)\delta(r)/r^{l+2}$ and $\frac{q^{2l+1}_\epsilon}{\tan[\eta_l(q_\epsilon)]}=(-\frac{1}{a_l}+r_l^e q^2_\epsilon+...)r_0^{-2l}$, and the expansion of $n_l(q_\epsilon r_{ij})$ at small $r_{ij}$, one could compute $\int_0^{r_0} Y_{lm}^\ast (\hat{\bf r}_{ij}) n_l^\ast(q_\epsilon r_{ij})\nabla_{ij}^2n_l(q_\epsilon r_{ij}) Y_{lm}(\hat{\bf r}_{ij})$ systematically. Not surprisingly, this removes the divergence in  $\int d{\bf k} \frac{\hbar^2k^2}{2M} \tilde{n}_{\bf k}$, since the divergence is indeed caused by the singular behavior of $n_l(q_\epsilon r_{ij})$ at small distance.

As a demonstration, here we show how to treat $l=1$, where we have
\begin{equation}
\begin{split}
&\frac{N(N-1)}{2}\frac{\hbar^2}{M}\sum_{m}\int d{\bf R}_{ij}  \int_0^{r_0}d{\bf r}_{ij}\Big\{  \Big[Y^*_{1m}(\hat{\bf r}_{ij})  \sum_{s=0} {\frac{\beta_{1s}g_{1m}^{*2s}}{r_{ij}^{2-2s}}}   \Big]\nabla_{ij}^2 \Big[   \sum_{s=0} {\frac{\beta_{1s}g_{lm}^{2s}}{r_{ij}^{2-2s}}}   Y_{1m}(\hat{\bf r}_{ij})\Big]\Big\}\\
\approx &\frac{N(N-1)}{2}\frac{\hbar^2}{M}\sum_{m}\int d{\bf R}_{ij}  \int_0^{r_0}d{\bf r}_{ij}\Big\{  Y^*_{1m}(\hat{\bf r}_{ij})\Big[ \frac{\beta_{10}g_{1m}^{*0}}{r_{ij}^{2}}+\beta_{11}g_{1m}^{*2}   \Big]\nabla_{ij}^2 \Big[   \frac{\beta_{10}g_{1m}^{0}}{r_{ij}^{2}} +\beta_{11}g_{1m}^{2}  \Big]Y_{1m}(\hat{\bf r}_{ij})\Big\}
\end{split}
\end{equation}
and $g_{lm}^{2s}= \int d\epsilon q_\epsilon^{2s} G_{lm}({\bf R}_{ij};\epsilon)$. 

The {{leading}} term gives
\begin{equation}
\begin{split}
&\frac{N(N-1)}{2}\frac{\hbar^2}{M}\sum_{m}\int d{\bf R}_{ij}  \int_0^{r_0}d{\bf r}_{ij}\Big\{  Y^*_{1m}(\hat{\bf r}_{ij}) \frac{\beta_{10}g_{1m}^{*0}}{r_{ij}^{2}} \nabla_{ij}^2  \frac{\beta_{10}g_{1m}^{0}}{r_{ij}^{2}} Y_{1m}(\hat{\bf r}_{ij})\Big\}\\
=&\sum_{m} \int \frac{d{\bf k}}{(2\pi)^3} \frac{\hbar^2 k^2}{2M}\Big\{ -\frac{C_{1m}}{k^{2}} Y^*_{1m}(\hat{\bf k}) Y_{1m}(\hat{\bf k})\Big\}
\end{split}
\end{equation}
The {{subleading}} term gives
\begin{equation}
\begin{split}
&\frac{N(N-1)}{2}\frac{\hbar^2}{M}\sum_{m}\int d{\bf R}_{ij}  \int_0^{r_0}d{\bf r}_{ij} Y^*_{1m}(\hat{\bf r}_{ij}) \Big\{  \frac{\beta_{10}g_{1m}^{*0}}{r_{ij}^{2}} \nabla_{ij}^2 \beta_{11}g_{1m}^{2} +\beta_{11}g_{1m}^{*2}  \nabla_{ij}^2  \frac{\beta_{10}g_{1m}^{0}}{r_{ij}^{2}} \Big\}Y_{1m}(\hat{\bf r}_{ij})\\
=&\sum_{m} \int \frac{d{\bf k}}{(2\pi)^3} \frac{\hbar^2 k^2}{2M}\Big\{ -\frac{C_{1m}^1}{k^{4}}f_1^1(kr_0) Y^*_{1m}(\hat{\bf k}) Y_{1m}(\hat{\bf k})\Big\}
\end{split}
\end{equation}
where $f_1^1(kr_0)=1-\frac{1}{2}j_0(kr_0)$.

Similarly for $l=2$,  one has
\begin{eqnarray}
&f_2^1(kr_0)=1-\frac{3}{2}\frac{j_1(kr_0)}{kr_0}\\
&f_2^2(kr_0)=1-\frac{C_{2m}^{20}}{C_{2m}^2}\left[ \frac{kr_0}{2}{j_1(kr_0)}+j_0(kr_0)\right]-\frac{C_{2m}^{11}}{C_{2m}^2}\left[3\frac{j_1(kr_0)}{kr_0}\right]\\
&C_{lm}^{ij}=(4\pi)^2{{N(N-1)}}\int d{\bf R}_{ij} { g_{lm}^{2i} g_{lm}^{*2j} }
\end{eqnarray}

For any $l$, there remains a term $\int_0^{r_0}Y^*_{lm}(\hat{\bf r}_{ij}) j_l^*(q_\epsilon r_{ij}) \nabla_{ij}^2 n_l(q_\epsilon r_{ij})Y_{lm}(\hat{\bf r}_{ij})$.

Again, by applying $\left(\frac{1}{r^2}\frac{d}{d r}r^2\frac{d}{dr}-\frac{l(l+1)}{r^2}\right) \frac{1}{r^{l+1}}=-(2l+1)\delta(r)/r^{l+2}$, we see that such a term leads to
\begin{equation}
\begin{split}
&\begin{split}
\frac{N(N-1)}{2}\frac{\hbar^2}{M}\sum_{lm}\int d{\bf R}_{ij}  \int_0^{r_0}d{\bf r}_{ij}\Big\{ & \Big[Y^*_{lm}(\hat{\bf r}_{ij}) \int d\epsilon G^*_{lm}({\bf R}_{ij};\epsilon) \left(\frac{q_\epsilon^{l+1}}{\tan[\eta_l(q_\epsilon)]}j_l(q_\epsilon r_{ij})\right)^\ast \Big]\\
 &\nabla_{ij}^2 \Big[ \int d\epsilon G_{lm}({\bf R}_{ij};\epsilon) \left(-q_\epsilon^{l+1} n_l(q_\epsilon r_{ij})\right)Y_{lm}(\hat{\bf r}_{ij})\Big]\Big\}
\end{split}\\
&=\sum_{lm}{ \frac{\hbar^2 r_0^{-2l}}{32\pi^2 M} \frac{C_{lm}}{a_l}-\frac{\hbar^2 r_0^{-2l}}{64\pi^2 M} r_l^e C_{lm}^1}
\end{split}
\end{equation}

The reason that only $C_{lm}$ and $C_{lm}^1$ show up here is because the low energy expansion of the phase shift $\frac{q^{2l+1}_\epsilon}{\tan[\eta_l(q_\epsilon)]}=(-\frac{1}{a_l}+r_l^e q^2_\epsilon+...)r_0^{-2l}$ has been used.

For the right hand side (RHS) of the equation (\ref{En}), since $\Psi$ is normalized to 1 and three-body physics leads to high order contributions in the dilute limit, we have
\begin{equation}\label{LHS}
\int' \prod_id{\bf r}_i | \Psi |^2=1-\frac{N(N-1)}{2}\sum_{lm}\int d{\bf R}_{ij} \int_0^{r_0} d{\bf r}_{ij} |\int d\epsilon\psi_{lm}({\bf r}_{ij};\epsilon) G_{lm}({\bf R}_{ij}, E-\epsilon)|^2\end{equation}

In low energy limit, we can expand the pair wave function $\psi_{lm}({\bf r}_{ij};\epsilon)$ as:
\begin{equation}
\psi_{lm}({\bf r}_{ij};\epsilon)\rightarrow \sum_s\psi^{(2s)}_{lm}({\bf r}_{ij})q^{2s}_\epsilon
\end{equation}
The equation (\ref{LHS}) can be expressed by using contact:
\begin{equation}
\begin{split}
P_{r_0}&=1-\frac{N(N-1)}{2}\sum_{lm}\int d{\bf R}_{ij} \int_0^{r_0} d{\bf r}_{ij} |\int d\epsilon \Big(\sum_s\psi^{(2s)}_{lm}({\bf r}_{ij})q^{2s}_\epsilon\Big) G_{lm}({\bf R}_{ij}, E-\epsilon)|^2\\
&=1-\sum_{lm}\Big[\frac{C_{lm}}{32\pi^2}\int_0^{r_0} |\psi^{(0)}_{lm}({\bf r}_{ij})|^2 d{\bf r}_{ij}+\frac{C^1_{lm}}{64\pi^2}\int_0^{r_0} \Big(\psi^{(2)\ast}_{lm}({\bf r}_{ij})\psi^{(0)}_{lm}({\bf r}_{ij})+\psi^{(0)\ast}_{lm}({\bf r}_{ij})\psi^{(2)}_{lm}({\bf r}_{ij})\Big) d{\bf r}_{ij}+\cdots\Big]
\end{split}
\end{equation}

All the integrals are purely two-body quantities, which can be obtained by either microscopic calculations or measurements in simple many-body systems where $C_{lm}$ are known.

\end{document}